*29 January 2003*

# Unexplained Sets of Seismographic Station Reports and A Set Consistent with a Quark Nugget Passage

David P. Anderson
Department of Geological Sciences
Southern Methodist University
dpa@io.isem.smu.edu

Eugene T. Herrin
Department of Geological Sciences
Southern Methodist University
herrin@passion.isem.smu.edu

Vigdor L. Teplitz
Department of Physics
Southern Methodist University
teplitz@mail.physics.smu.edu

Ileana M. Tibuleac
Weston Geophysical
Boston, MA
ileana@westongeophysical.com

ABSTRACT

In 1984 Edward Witten proposed that an extremely dense form of matter composed of up, down, and strange quarks may be stable at zero pressure (Witten, 1984). Massive nuggets of such dense matter, if they exist, may pass through the Earth and be detectable by the seismic signals they generate (de Rujula and Glashow, 1984). With this motivation we investigated over 1 million seismic data reports to the U.S. Geological Survey for the years 1990-1993 not associated with epicentral sources. We report two results: (1) with an average of about 0.16 unassociated reports per minute after data cuts, we found a significant excess over statistical expectation for sets with ten or more reports in ten minutes; and (2) in spite of a very small a priori probability from random reports, we found one set of reports with arrival times and other features appropriate to signals from an epilinear source. This event has the properties predicted for the passage of a nugget of strange quark matter (SQM) through the earth, although there is no direct confirmation from other phenomenologies.

## I. INTRODUCTION

We present evidence for detection of a 1993 seismic event with an epilinear source. We are aware of only one model that predicts seismic line events with a frequency on the order of one a year, namely the passage of "nuggets" of quark (or gluon) matter through the earth.

In 1984, Witten pointed out that, while matter made of up and down quarks is not stable, because ups and downs condense to form protons and neutrons, matter made of up, down, and strange quarks, SQM, may well be stable (Witten, 1984). This is because of the roughly 10% decrease in kinetic energy from having three Fermi seas with which to satisfy the Pauli principle instead of just two (with the same potential energy). Witten also suggested a scenario for early universe SQM nugget production, variations of which are still under debate (Cottingham, Kalafatis and Mau, 1994, but see Cho et al, 1994), as well as the possibility of strange quark nuggets (SQN) as dark matter candidates. An informative non-mathematical discussion of SQM can be found in Siegfried, (2002).

SQN's would not be limited in total baryon number (Farhi and Jaffe, 1984) as is ordinary matter. Thus massive nuggets of SQM are possible. They would have nuclear densities (~ $10^{14}$ gm/cm$^3$). Because of the larger mass of the strange quark, an SQN with 3A quarks would have an excess of positively charged quarks over strange, negatively charged quarks, and hence a net positive charge, from "nuclear particles," which would be balanced by an electron cloud. For $M>10^{-9}$ gram, the cloud would be mostly inside the nuclear part of the SQN. The SQN would be nearly neutral and, with high mass and low abundance, would not interact appreciably with electromagnetic energy; hence its suitability as a dark matter candidate. Finally, deRujula and Glashow (1984) discussed seismic detection of a massive SQN passage through the earth.

Recently, NASA's Chandra X-ray Satellite observed two neutron stars, one of which appeared too small (Drake, 2002) and one of which appeared too cold (Slane, 2002) to fit the standard model of neutron stars (Shapiro and Teukolsky, 1983). These observations could be consistent with the stars being composed, at least in part, of strange quark matter. However the observations are subject to uncertainties (Walter and Hefland, 2002) and, even should those stars be composed in part of strange quark matter under pressure, it would not necessarily follow that strange quark matter would be stable under zero pressure. Thus it is still a matter of debate as to whether SQN's exist. It should be noted that there exist other quark (and gluon) models that would for seismological purposes be indistinguishable from the SQM model discussed here (for example, Zhitnitsky, 2002).

While strange quark matter motivated our work, in the present paper we concentrate on the seismic analysis and leave questions of interpretation for future publications. We also note that while our candidate is, we believe, very strong, firm conclusions from it await confirmation from further seismological analysis and application of other phenomenology. Finally, we cannot rule out the possibility that the origin of the set of reports that constitute the candidate is related to that of the other sets of reports in a significant excess over statistical expectations of 10 or more random arrival times within a ten minute window.

Section II reviews previous Monte Carlo work undertaken to determine the feasibility of the analysis of this paper. Section III describes the data used herein, and the cuts to the data made to decrease backgrounds. Section IV describes the search for instances of a sufficient number of unassociated reports in a time window to identify an epilinear event. It reviews frequencies of such sets of reports, expected on the basis of a random distribution, for varying numbers of reports and varying time windows, given the actual frequency of unassociated reports. It finds 17 such sets of ten or more reports within 10 minutes for 1990-1993. About one would be expected from a random distribution. Section V describes the search for a fit to an epilinear event performed for each of the 17. Section VI addresses the case in which an excellent fit was obtained from the set of first signal arrival times and supporting evidence from available waveform data, including from arrays which give valuable pointing. Section VII presents a brief summary and conclusions. Finally, it should be noted that, although the work herein was motivated by the possibility of seismological indications of SQN passage events, the results of Section IV indicate a potentially important question for seismology itself.

## II. Review of previous work

Two of us examined detection of SQN seismic signals via a Monte Carlo calculation (Herrin and Teplitz, 1996). Briefly, a multi-ton sized SQN would have dimensions of tens of microns, the size of red blood cells. As it passed through the earth it would break inter- and intra-molecular bonds, like a stone dropped in water, producing a seismic signal. The rate of seismic energy [E] production would be given by

$$dE/dt = f\,\alpha\rho V^3$$

where $\alpha$ is the SQN cross section, $\rho$ is the nominal earth density, V is the SQN speed (on the order of a few hundred km/sec, the galactic viral velocity), and f is the fraction of SQN energy loss that results in seismic waves rather than other dissipation such as heat or breaking rock. Underground nuclear explosions have f of about 0.01, chemical ones about 0.02. The small size of SQN, which enhances coherence, depresses random motion and yields a high ratio of surface area to energy generating volume, implies that f might be larger for the SQN case.

A Monte Carlo method was used to identify the extent to which nuggets of stable strange quark matter (SQNs), should they in fact exist and have densities in the $10^{14}$ gm/cm$^3$ range as expected, could be detected seismically. An isotropic, Maxwellian galactic distribution was assumed and account taken of the Sun's velocity with respect to the galactic center of mass. A model was used with 287 actual seismographic stations, 48 of which have sufficient sensitivity to detect 1 kT of TNT with 1% coupling at 5000 km. A single average global sound propagation speed of 10 km/s was used. A 5% (f = 0.05) coupling to seismic waves for SQNs was assumed.

An SQN event should have a distinctive signal because of the large ratio (30:1) of SQN speed to speed of sound in the Earth. Detection of an SQN passage would require at least six stations to fix its impact time and location and its (vector) velocity. Seismic detection of signals by at least seven stations was required in order to separate SQN events from random spurious coincidences (Fig. 1). This is because 6 random arrival times might possibly fit the 6 parameters needed, but getting the same fit from all subsets of 6 drawn from seven or more would be very unlikely. 120,000 random geometries were generated. For about a twelfth of the geometries, SQNs with masses of or below one metric ton could be detected in the simplified earth model used. For about a third of those geometries, nuggets of or below ten metric tons could be detected.

The Monte Carlo study served as a guide for the present work. It showed that almost all (98.5%) of the detections of passages of SQNs of minimally detectable mass would be by the 48 "Class 1" seismographic stations sensitive to 1 Hz waves of energy density 0.133 gm/cm$^2$ sec or better, corresponding to the capability to detect a well coupled underground nuclear explosion of 1 kT at 5000km. In the present study we search for seismic line events which might result from an SQN passage. Station observations of such an event would not be associated by current methods that assume a point source for all small seismic events.

## III. The Data

Data were collected from the United States Geological Survey archive from 02 February 1981 through 31 December 1993, in USGS READING (RDG) report format. These data consist of reports turned in to the USGS from seismographic stations around the world largely at the discretion of individual human analysts, prior to the widespread adoption of automated reporting methods in the mid-1990s. Most of these reports have therefore been subjected to a kind of pre-screening based on the experience and judgment of the station analysts, who are less likely than automated systems to report signals from cultural sources, meteorological events, or spurious equipment noise.

---------------------------------------------------------------------------

The USGS Reading (RDG) Data Set
02 Feb. 1981 through 31 Dec. 1993

| | |
|---|---|
| Duration of data: | 13 years |
| Total reports: | 9,128,892 |
| Total associated: | 5,889,684 |
| Total unassociated: | 3,239,208 |

---------------------------------------------------------------------------

Approximately 8000 stations are included in the database. The base includes over 9 million separate reports, about 6 million of which have been previously associated with epicentral locations and more than 3 million of which remain unassociated. It is among the unassociated reports that we expect to find the signature of linear seismic sources, should they exist.

Model of P-wave arrival timing for supersonic (~250 km/sec) linear source

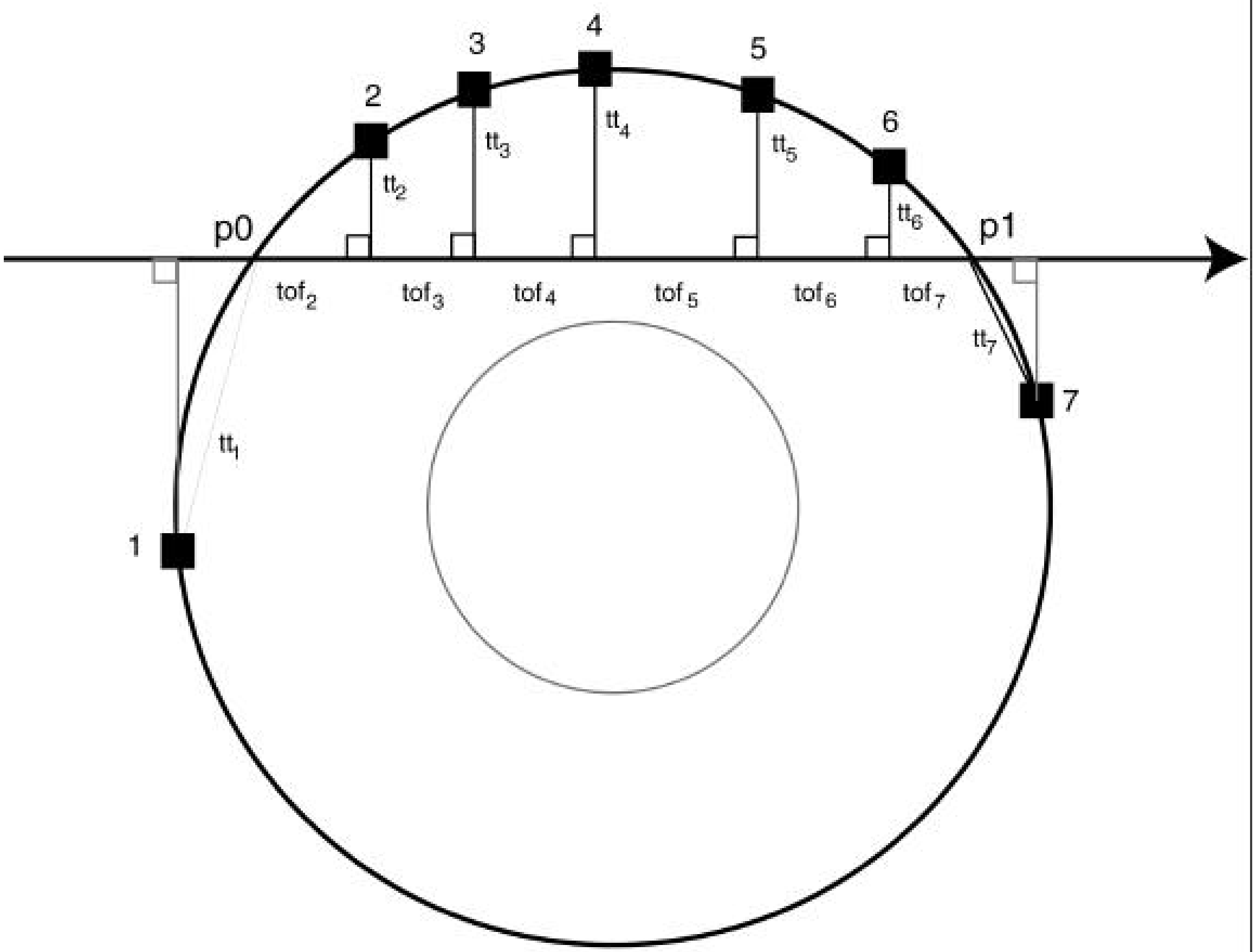

The linear source is determined by 6 parameters:

p0 = entry point in latitude, longitude
p1 = exit point in latitude, longitude
$t_0$ = source (entry) time
v = source speed in km/s

Arrival time for a station is given as:

arrival = $t_0$ + tof + tt

where:

tof = time of flight from entry to point of closest approach (POCA)
= distance from entry to POCA / v
tt = IASPI travel time from POCA to station

The first arrival at a station is expected to come from the point of closest approach between the station and a line defined by the entry and exit points of the linear source. This POCA will be slightly offset from true perpendicular by the angle of the MACH cone of the supersonic source. At the expected velocities, ~250 km/sec, this offset should be negligible. A special case arises when the POCA is outside the sphere of the earth, ilustrated here by stations 1 and 7. In this case the expected first arrival for station 1 is just $t_0$ + the IASPI travel time from P0 to the station. For station 7 the first arrival is $t_0$ + total tof + IASPI travel time from P1 to the station. The order of first arrivals for the example illustrated would be 6,2,7,3,5,4, and 1.

**Figure 1.**

Initial inspection of the unassociated data revealed the presence of large numbers of reports that could be associated as core phases of large earthquakes ( > 4.5 Mg), often with travel times in excess of 30 minutes. These late arrivals make automated association with the source earthquake problematic, hence their prominence in the unassociated data set.

In order to remove these reports, the USGS Preliminary Determination of Epicenters (PDE) database was obtained. The data available electronically overlap the RDG data in the years 1990 through 1993. Using the PDE data all reports within 60 minutes following any magnitude 4.5 or greater epicenter determination were removed. This process filtered out about half of the reports.

The remaining unassociated reports for 1990-1993 were further filtered to remove all reports except those from the 48 most sensitive "Class I" seismographic stations, based on the earlier study (Herrin and Teplitz, 1996). This left about 15% of the original unassociated RDG reports. Statistics for 1993 are typical.

---

1993 unassociated RDG reports:

| | | |
|---|---|---|
| Total reports: | 284,809 | |
| After PDE filtering: | 152,272 | (53%) |
| After station filtering: | 54,101 | (18%) |
| | | |
| Total minutes in 1993: | 525600 | |
| Total minutes removed: | 191146 | (36%) |
| Total minutes remaining: | 334454 | (64%) |

---

For 1993 the data set contains about fifty-four thousand unassociated reports submitted by human analysts from the 48 best seismographic stations all over the world, which are not late arrivals from any recognized large earthquakes, and which had not as of 1993 been associated with any traditional seismic event. A further possible reduction, elimination of reports subsequently associated by others, in particular by the International Seismic Center (ISC), was not made until after searching for candidate sets of reports in the data base as described above.

Seismographic stations in the Northern Hemisphere, especially North America, are under-represented in this set. These stations do not routinely submit reports that are not already associated with a particular seismic event.

The completed filtered data set of unassociated reports from the years 1990-1993 was then searched for seismic arrival times consistent with epilinear and epicentral sources.

## IV. The Search

### A. Travel times

Figure 1 shows the geometry and arrival timing for a hypothetical linear source. Signals that pass through the earth's core are quite complex due to reflections and refractions at the boundaries, and hence were not considered for this study. The fact that the earth's core is roughly half an earth radius implies that about 75% of randomly oriented linear sources will not pass through the core.

Seismic travel times through the earth are well known (IASPEI, 1991) and calibrated down to point source depths of 700 kilometers. For this study additional travel time tables were generated by ray tracing through the standard earth model (Kennett, 1991) to a depth of 2880 kilometers: the core-mantle boundary. The travel travel time tables thus generated were compared to published data (IASPEI, 1991) down to 700 kilometers depth and were also compared with signals reflected from the core-mantle interface, and are in good agreement with both. This extended depth travel time table is available for download from http://www.geology.smu.edu.

### B. Arrival windows

Our initial search looked for $K \geq 10$ unique station reports in an event window of $T \leq 600$ seconds length. The 600 second arrival time window chosen for this study covers the range of expected travel times for a linear source of 2880 kilometers maximum depth for stations at 5000 kilometers maximum distance (Herrin and Teplitz, 1996).

For random reports uniformly distributed the expected number of reports (K) in a window of T minutes is Poisson distributed. A calculation based on this distribution for the 54101 unassociated reports from 1993 is shown in Figure 2 and summarized in Table 1. A false hit is the occurrence of K or more random reports in an interval of T minutes. Note that 50,000 reports in about 33,000 minutes gives a rate of about 0.16 reports per minute. The upper line plots the rate of K or more station reports occurring in a 600 second time window. The expected rate for 10 or more randomly distributed reports is 0.26, or about once every four years. The actual data often includes multiple reports from the same station in one event window, so more than 10 reports are required to satisfy the $K \geq 10$ requirement. From Table 1 the rate for 12 or more random reports in a 600 second window is about .005, on the order of once every 200 years.

The candidate set of reports for which we have fit a linear source in November 1993 consists of 9 station arrivals, 7 Class I stations and 2 others, which span an actual time window of about 130 seconds. The lower line in Figure 2 plots the expected occurrence of K or more random reports in a 2.2 minute window. From Table 1 the expected rate for 7 randomly distributed Class I station reports in 2.2 minutes is about .016 per year.

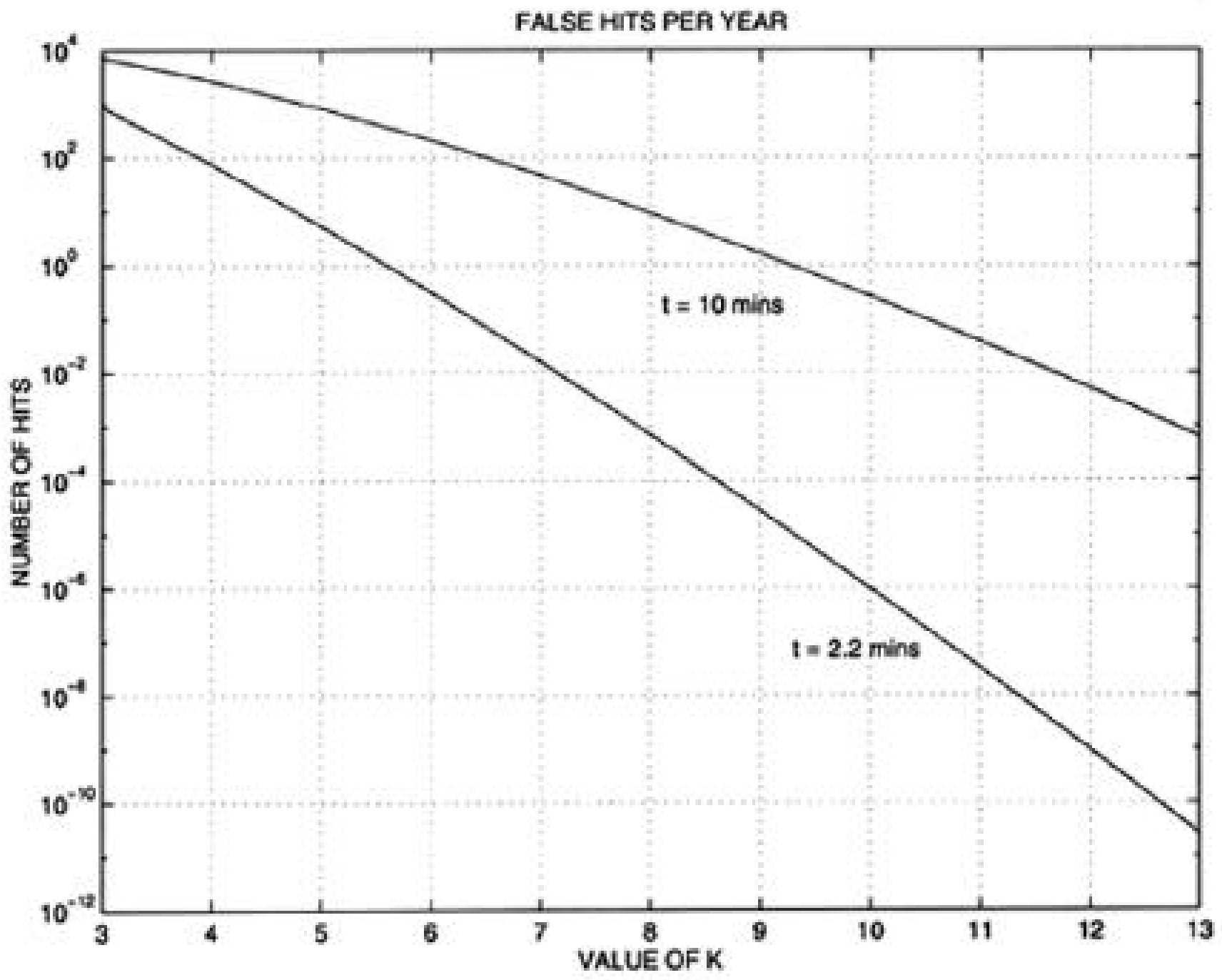

**Figure 2**

False Hits Per Year

| T= | 2.2 mins | 10 mins |
|---|---|---|
| K=3 | 8.7715e+2 | 7.4016e+3 |
| K=4 | 7.6621e+1 | 2.7196e+3 |
| K=5 | 5.3885e+0 | 8.2570e+2 |
| K=6 | 3.1692e-1 | 2.1284e+2 |
| K=7 | 1.6061e-2 | 4.7578e+1 |
| K=8 | 7.0870e-4 | 9.3774e+0 |
| K=9 | 2.7916e-5 | 1.6518e+0 |
| K=10 | 9.9042e-7 | 2.6290e-1 |
| K=11 | 3.1961e-8 | 3.8102e-2 |
| K=12 | 9.4587e-10 | 5.1011e-3 |

Table 1.

C. Reports per window

Table 2 shows the actual data rate for the 54101 unassociated station reports of 1993 as returned by the association algorithm used for the linear event search. It is apparent that the number of occurrences of K or more reports in a window of T seconds is significantly larger than expected for random reports. For the T=600 second window, the search found 18 occurrences of 10 or more stations. Comparing these 18 sets of associations with published data (International Seismological Center, 2002) we were able to find one or more station reports in 7 of the sets that have been associated with eipcentral locations calculated subsequent to 1993. The remaining 11 sets of 10 or more reports consist of arrivals that remain unassociated and are not overlapped in time with any known epicentral locations the authors have been able to ascertain. This is a rate about 30 times that expected for random coincidence.

Applying this same process to the years for which we have electronic versions of the station reports and the PDE, we were able to identify 6 additional candidate sets of reports which fit our search criteria ($K \geq 10$ and $T \leq 600$) for a total of 17 sets for the years 1990-1993. Using the earthquake location program HYPOSAT (Schweitzer, 1997) we identified one set of reports in October 1993 as associated with a small earthquake, probably mixed with reports of a second, unrecognized, earthquake. None of the other sets of reports could be made to fit epicentral sources, but a set was found from November 1993 that is a very good fit to an epilinear source. This paper concentrates on that candidate set of reports. More generally, however, we note that the ratios of the entries in the 600 column of Table 2 to the expected numbers from Table 1 increase with K, leading us to reject the hypotheses that the reports are random. We plan to study such features of unassociated reports in detail in follow-on work.

1993 truncated unassociated data, 54101 total reports

| T = | 240 | 360 | 480 | 600 | 720 | 840 | 960 |
|---|---|---|---|---|---|---|---|
| K = 2 | 13138 | 13850 | 13839 | 13542 | 13026 | 12419 | 11858 |
| K = 3 | 4476 | 5968 | 6923 | 7498 | 7817 | 8010 | 8029 |
| K = 4 | 1396 | 2217 | 2947 | 3574 | 4089 | 4515 | 4837 |
| K = 5 | 422 | 785 | 1171 | 1573 | 1939 | 2303 | 2614 |
| K = 6 | 118 | 281 | 462 | 661 | 876 | 1072 | 1313 |
| K = 7 | 32 | 88 | 165 | 251 | 365 | 484 | 624 |
| K = 8 | 11 | 28 | 56 | 95 | 145 | 208 | 272 |
| K = 9 | 3 | 11 | 26 | 44 | 63 | 86 | 122 |
| K = 10 | 1 | 4 | 11 | 18 | 26 | 36 | 51 |
| K = 11 | 0 | 0 | 3 | 10 | 13 | 21 | 26 |
| K = 12 | 0 | 0 | 0 | 1 | 3 | 7 | 13 |
| K = 13 | 0 | 0 | 0 | 0 | 1 | 1 | 2 |
| K = 15 | 0 | 0 | 0 | 0 | 0 | 0 | 0 |

**Table 2**

## V. The Linear Fit.

Groups of 10 or more station reports in a 600 second window returned by the first stage of the linear association algorithm for the years 1990-1993 were subjected to two forms of analysis. The earthquake location program HYPOSAT (Schweitzer, 1997) was used to attempt to find an epicentral source, and the linear event recognizer LEVR, written for this study, was used to attempt to find a linear source. Both algorithms return a figure of merit that is the total RMS of the travel time residuals for the observed arrival time differences minus the predicted arrival time differences for the proposed source. The method for determining this RMS value used by HYPOSAT is extensively documented (Schweitzer, 1997). The method used by the LEVR algorithm is summarized as follows.

### A. The Point of Closest Approach.

Seismic theory for signals with a linear source moving at supersonic speed through the earth implies that the first arrival at a given seismic station will come, to a good approximation, from the Point of Closest Approach (POCA) of the source to the station. The POCA is the intersection of the trajectory with a perpendicular from the station to the line defined by the entry and exit points of the linear source, as diagrammed in Figure 1.

The first arrival signal source is offset slightly from this intersection (POCA) by the angle of the MACH cone produced by the supersonic transit. In practice, the effect of the cone angle is negligible at the expected velocities and can be ignored. The galactic viral velocity for trapped materials is about 250 km/s. In cases where the point of closest approach lies outside of the Earth, the first arrival signal source is the closer of the entry or the exit point.

Referring to Figure 1, the arrival time at any station is determined as the sum of the source entry time, the time-of-flight from entry to POCA for that station, and the seismic travel time from the POCA to the station. The model assumes a spherical earth. Two additional second order corrections were manually included in the final fitting of the event of November 1993: an elevation correction for the individual seismic stations, and a source correction for signals originating in oceanic rather than continental crust.

### B. The Test of Hypotheses

The test for fit to a linear source consists of constructing a hypothetical source as defined by an entry, exit, and speed. The POCA for each station in the event window is calculated and the POCA-to-station travel times are obtained. A range of speeds from 100 to 800 km/sec in 50 km/sec steps is used to calculate the time-of-flight, and a set of predicted arrival times is determined.

Each station is then taken in turn as the reference station, and its arrival time subtracted from the arrival times of all the others to produce an array of travel time differences, $\Delta t$.

A similar $\Delta t$ array is created for the observed arrivals using the same reference stations, and the two arrays are differenced to produce an array of residuals. Next the residuals are sorted and the lowest N residuals are used to calculate the total RMS error, where N is an input parameter $\geq 7$. The residual, source geometry, velocity, and reference station are returned for the linear source with the lowest RMS.

LEVR repeats this process for all "acceptable" linear sources with entry and exit points located on a global 0.1x0.1 degree grid. (A linear source was not considered "acceptable" for a given geometry for a set of reports if the signal received by a station from its POCA would require passage of the signal through the Earth's core, i.e. if there is core shadowing.) To make this process tractable with the computing resources available, the search is performed using a collapsing grid technique. A global search space of 360x180 degrees is first executed on a 10x10 degree grid. That is, each end of the source is moved through 360 degrees of longitude and 180 degrees of latitude in 10 degree steps. The best fit from that stage becomes the input for the next stage, in which a search space of 20x20 degrees is executed on a 1x1 degree grid. Finally a 2x2 degree search is executed on a 0.1x0.1 degree grid, and the velocity is varied in 10 km/sec steps. The RMS figure of merit returned by the last stage is the best fit for a linear source to the given set of station reports. That number is typically on the order of 100 seconds for a set of random reports. For a synthetic linear source it is typically less than .01 seconds.

### C. Linear Fits to Synthetic Data

A synthetic linear source with an entry at 45,0 (lat,lon) an exit at 45,180, and a velocity of 400 km/sec was generated for the 48 Class I seismic stations. Each of 10 tests of random subsets of combinations of 8 stations returned the correct strike geometry and velocity with an average total residual RMS of 0.002s.

A second source was generated for the same geometry and Gaussian noise with a mean of 0 and a standard deviation of 0.3 seconds was added to the synthetic arrival times. Ten tests of random subsets of combinations of 8 stations each returned the correct source geometry and velocity with an average total residual RMS of 0.3s.

A third linear source was generated using the noisy reports from the second test mixed with 3 random arrivals for a total of 11 reports in each window. Each of ten tests of random subsets of combinations of 8 stations was able to correctly associate the stations with the linear source and reject the 3 random arrivals.

### D. Linear Fits to Random Arrivals

A Monte Carlo simulation of 1000 sets of 9 non-duplicate reports of station arrivals randomly distributed in a 600 second window was generated for randomly chosen subsets of the 48 Class I stations, weighted according to their frequency of reporting for the year 1993. The number 9 was chosen for comparison to the reports found in November 1993, which returned an excellent fit (0.23 seconds RMS) for 9 stations. These 1000 windows were then tested for fit to a linear source, the results sorted in ascending order of total residual RMS, and plotted as the upper line in Figure 3.   The lowest figure of merit returned by these tests was 33.3 seconds RMS.

The candidate set of 9 reports from November 1993 spans a time window of 130 seconds, rather than the full 600 seconds, and these simulations were re-run with 2.2 minute windows accordingly. The lowest figure of merit returned by 1000 fits was 9.7 seconds RMS, as shown in the lower plot in Figure 3 -- ten times larger than our search criteria for a linear source.

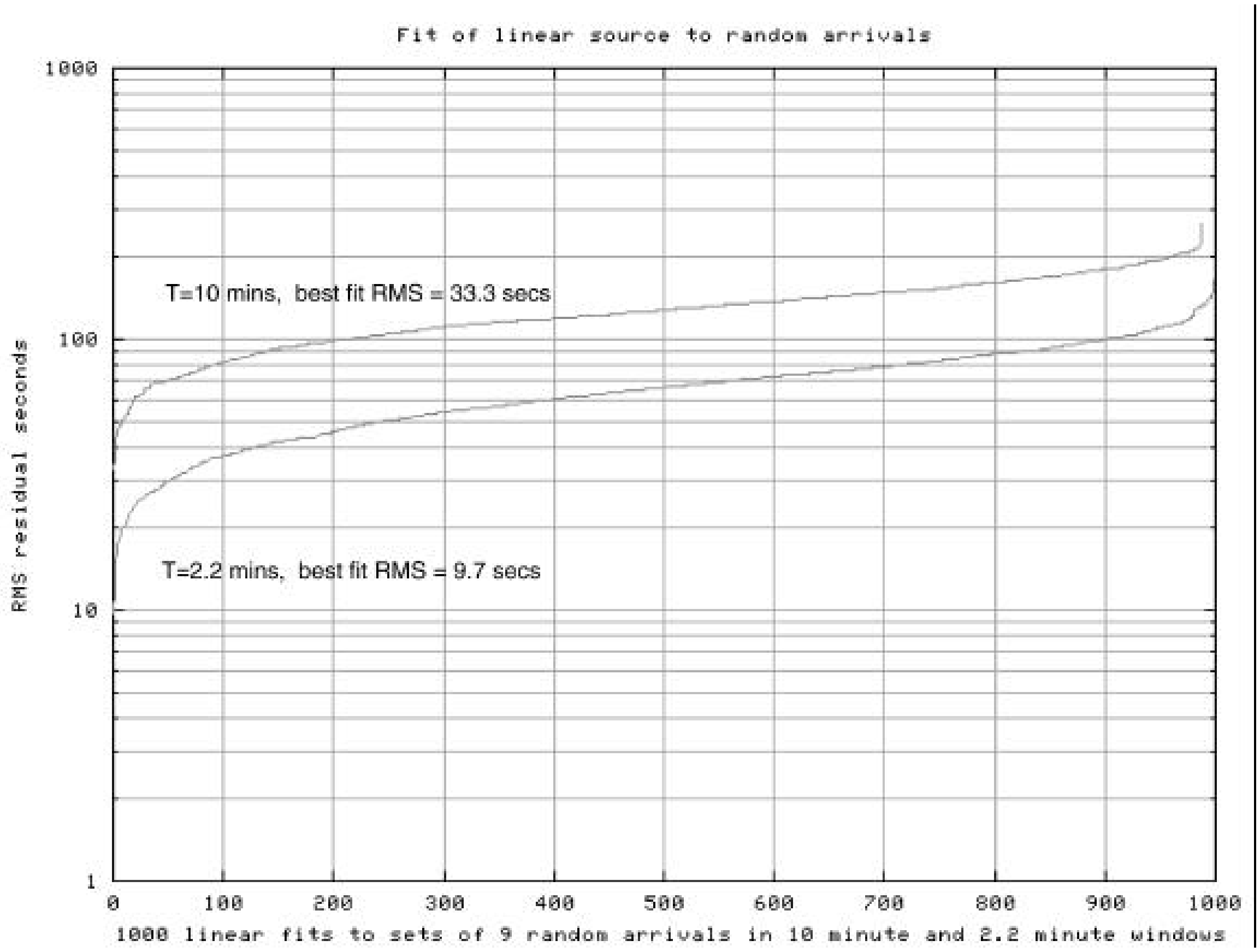

**Figure 3**

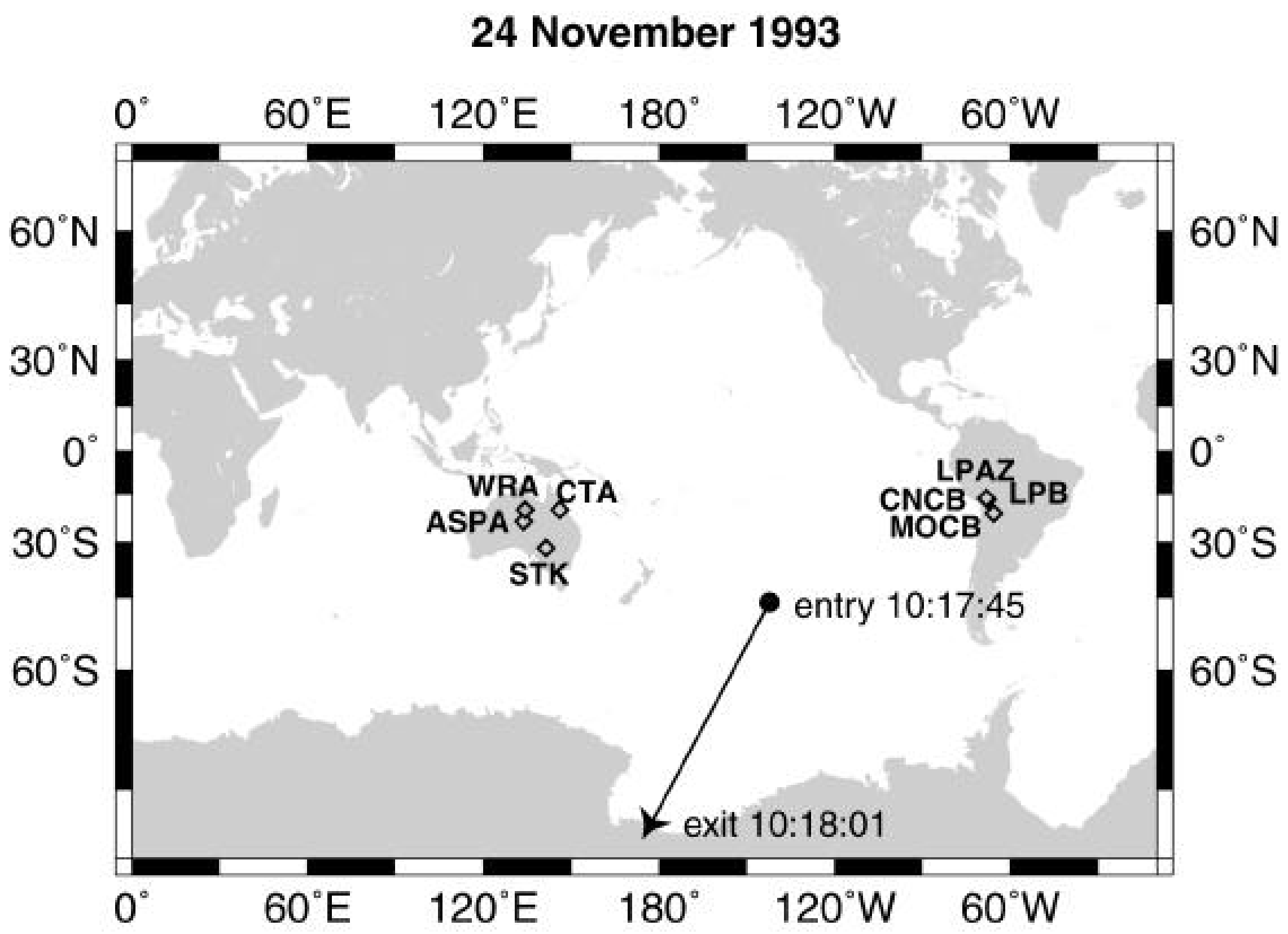

**Figure 4. Surface trace for November 1993 linear event.**

VI. The November 1993 Event

A. Fit to an epilinear source

Figure 4 plots the surface trace for a linear source fit of the 9 unassociated station reports from November 24, 1993. The source has an entry time of 10:17:45 in the South Pacific and an exit point 16 seconds later in the Ross Ice Shelf near the South Pole. The model and residuals are given in Table 3. All residuals are less than ± 0.4 sec., an excellent fit of data to the model. The root-mean-square (RMS) residual is 0.23 sec. This result is better than many obtained for well-located earthquakes.

The linear event search for November 24,1993, originally returned 7 stations from a set of 11 as fitting an eiplinear source. Two of the reports in the window were subsequently identified as belonging to a small local event, and two others were removed as duplicates. Thereafter the search criteria were expanded to include stations not in the 48 Class I set, and two more reports were found in the event window: WB2 and MOCB. These also fit the linear source. The geometry of the source suggests that stations at the South Pole and New Zealand might also have seen the signal, and waveforms were obtained for these stations. No signals were found, and it was determined that the expected signal amplitudes were below the background noise levels in both cases.

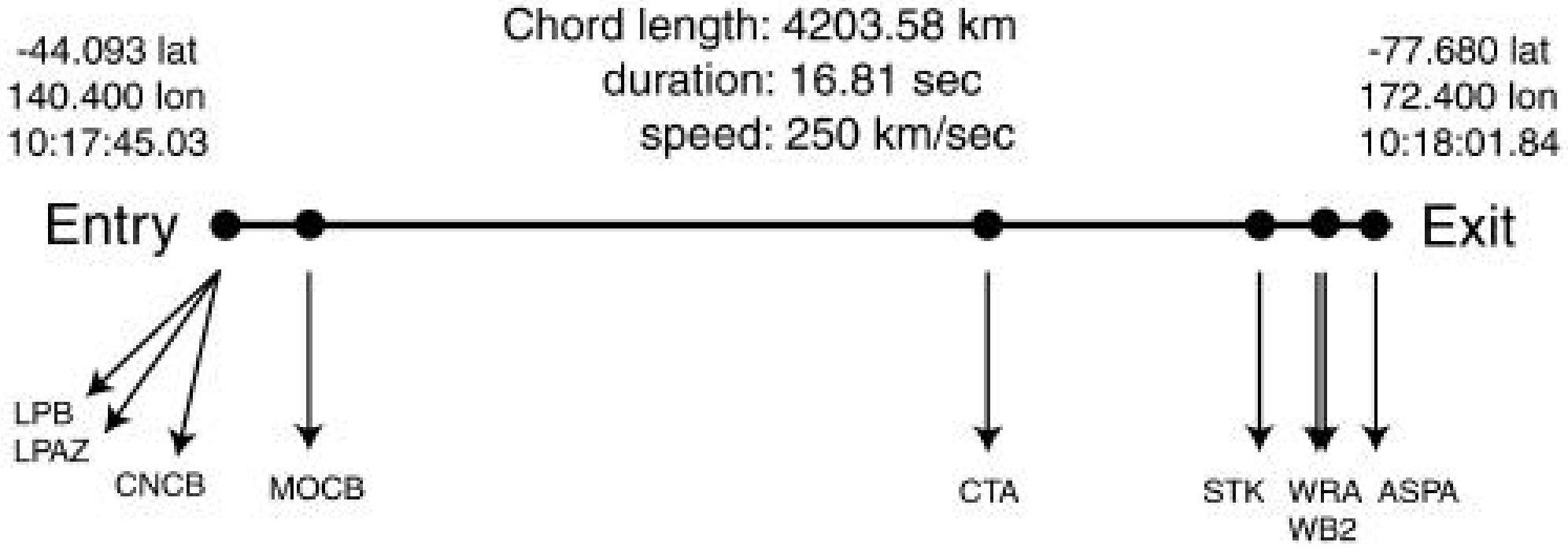

**24 Nov 1993 linear source: points of closest approach**

| station | distance from station (deg) | depth (km) | IASP travel time (sec) | elevation corr(sec) | source corr(sec) | linear distance from entry (km) | time of flight from entry (sec) |
|---|---|---|---|---|---|---|---|
| STK | 48.229 | 114.704 | 508.87 | .04 | 0.00 | 3824.511 | 15.298 |
| CTA | 59.120 | 302.446 | 569.15 | .07 | 0.00 | 2910.910 | 11.643 |
| ASPA | 57.049 | 0.128 | 587.66 | .11 | 0.00 | 4202.949 | 16.812 |
| WRA | 60.919 | 56.830 | 606.83 | .07 | 0.00 | 4023.924 | 16.095 |
| WB2 | 60.918 | 57.156 | 606.79 | .07 | 0.00 | 4023.286 | 16.093 |
| MOCB | 65.273 | 108.825 | 630.03 | .71 | 0.00 | 357.413 | 1.430 |
| CNCB | 65.933 | 0.000 | 647.82 | .86 | 0.82 | 0.000 | 0.000 |
| LPB | 66.033 | 0.000 | 648.46 | .66 | 0.82 | 0.000 | 0.000 |
| LPAZ | 66.171 | 0.000 | 649.35 | .95 | 0.82 | 0.000 | 0.000 |

| Station | observation | - ( source_time | + poca_tof | + ttime | + elev | - source ) | = | residual |
|---|---|---|---|---|---|---|---|---|
| STK | 10:26:29.30 | - (10:17:45.03 | + 15.298 | + 508.87 | + .04 | - 0.00 ) | = | 0.060 |
| CTA | 10:27:26.00 | - (10:17:45.03 | + 11.643 | + 569.15 | + .07 | - 0.00 ) | = | 0.104 |
| ASPA | 10:27:49.90 | - (10:17:45.03 | + 16.812 | + 587.66 | + .11 | - 0.00 ) | = | 0.286 |
| WRA | 10:28:07.60 | - (10:17:45.03 | + 16.095 | + 606.83 | + .07 | - 0.00 ) | = | -0.427 |
| WB2 | 10:28:08.00 | - (10:17:45.03 | + 16.093 | + 606.79 | + .07 | - 0.00 ) | = | 0.015 |
| MOCB | 10:28:17.20 | - (10:17:45.03 | + 1.430 | + 630.03 | + .71 | - 0.00 ) | = | -0.002 |
| CNCB | 10:28:33.00 | - (10:17:45.03 | + 0.000 | + 647.82 | + .86 | - 0.82 ) | = | 0.108 |
| LPB | 10:28:33.00 | - (10:17:45.03 | + 0.000 | + 648.46 | + .66 | - 0.82 ) | = | -0.332 |
| LPAZ | 10:28:34.70 | - (10:17:45.03 | + 0.000 | + 649.35 | + .95 | - 0.82 ) | = | 0.188 |

Total RMS residual **0.234** sec

**Table 3**

B. Fit to an epicentral source

Using the earthquake location program HYPOSAT (Schweitzer, 1997) we tried to fit the nine arrival times, back azimuth and slowness (the reciprocal of horizontal phase velocity) values from the two Australian arrays (WRA and ASPA) with a point source model. The iterative location process did not converge until depth was fixed at 109 km, a very unlikely depth for an earthquake near the Pacific-Antarctic Ridge. After 158 iterations the program produced a final location with an RMS residual of 2.27 sec, about 10 times the linear fit. However, HYPOSAT was unable to fit the point source without rejecting the arrival at STK in Australia, where the arrival was clear and well timed, and without large residuals for ASPA and WRA. With the depth manually fixed at 15km, HYPOSAT was able to return a location with an RMS residual of 2.13 seconds.

We attempted to fit a point source to the 5 Australian stations without including the South American station reports. This was more successful, and with a depth fixed at 15km HYPOSAT returned a location with an RMS residual of .776 seconds without rejecting the STK report. The problem with this scenario is that for the calculated magnitude of this event, about 4.2, the South American stations would have certainly seen the same event. If there were two overlapping events, both sets of stations should have seen both sets of events. Examination of the waveforms rules this out. Both sets of stations saw only a single event.

We consider the point source model to be highly unlikely because its RMS residual was 10 times larger than the RMS value obtained with the linear source model.

C. Waveforms

Referring to the surface trace of the linear source model for this event shown in Figure 4 and Table 3, the POCAs for the Australian stations are along the last third of the linear path. The POCA's for the Bolivian stations were at or very near the entry point. Back-azimuths computed for the two Australian arrays (ASPA and WRA) point to the POCA's.

Waveforms for these two arrays are shown in Figs. 5 and 6. We expect the first arrival to be essentially from the POCA followed by energy arriving from the linear source away from the POCA in both directions. The waveforms seen at the two arrays are very unusual for an event of magnitude between 4 and 4.5 as determined from the amplitude of all arrivals. There is a sharp on-set with nearly constant amplitude for several seconds after which the amplitude decreases markedly. However, the signal continues above ambient noise for over one minute. The time at which the amplitude decreases is the predicted time that the source leaves the earth for three of the Australian stations. After that time only energy from the path between the entry and the POCA would be seen.

Table 4 shows the length of the high amplitude part of the signals seen at the four Australian stations compared with the seismic travel time (TT) differences for the POCA and exit.

| Station | TT to POCA | TT to Exit | ΔTT | TOF | Total | Duration of High amplitude |
|---|---|---|---|---|---|---|
| ASPA | 592.6 | 592.5 | 0.5 | 0 | 0.5 | 6 |
| WRA | 612.1 | 617.3 | 5.2 | 0.9 | 6 | 8 |
| STK | 514.0 | 524.7 | 10.7 | 1.9 | 13 | 12 |
| CTA | 575.0 | 605.5 | 30.5 | 5.2 | 36 | 37 |

**Table 4**

The comparisons are consistent with the previously stated explanation for the shape of the waveforms for three of the stations; however, the high amplitude segment for ASPA is significantly longer than expected. For three of the stations the amplitude is 3 to 5 nanometers with a dominant period of less than 1 second. For ASPA the amplitude is 16 nm with a period of 1.3 seconds. The POCA for ASPA is in the Ross Ice Shelf. It is possible that resonances in the ice and underlying ocean could have led to the anomalously large amplitude at ASPA. It is not possible to accurately model this effect because the ice thickness is poorly known, but a resonance of 1.3 seconds is possible within the limits we can assign to the thickness of the ice and water layers.

Waveforms from La Paz (Figure 9) do not show this pattern but persist at essentially the initial amplitude for at least one minute. This pattern is consistent with a source moving away from the station with a POCA at or near the point of entry.

The data for this event are consistent with the linear source model in the arrival times, source direction as determined from the arrays, and waveform patterns. The data cannot be forced to fit a point source model without unreasonably large errors. The waveforms do not show the patterns expected for small earthquakes.

Finally, we note that detection of one event in a four year period of passage of a multi-ton SQN through the earth would appear to be consistent with likely possible frequencies (Herrin and Teplitz, 1996). Quantitative comparisons, however, await detailed consideration (in progress) of the fraction of energy loss that goes into seismic waves for such an event, the effective acceptance of our search procedure, as well as astrophysical inputs.

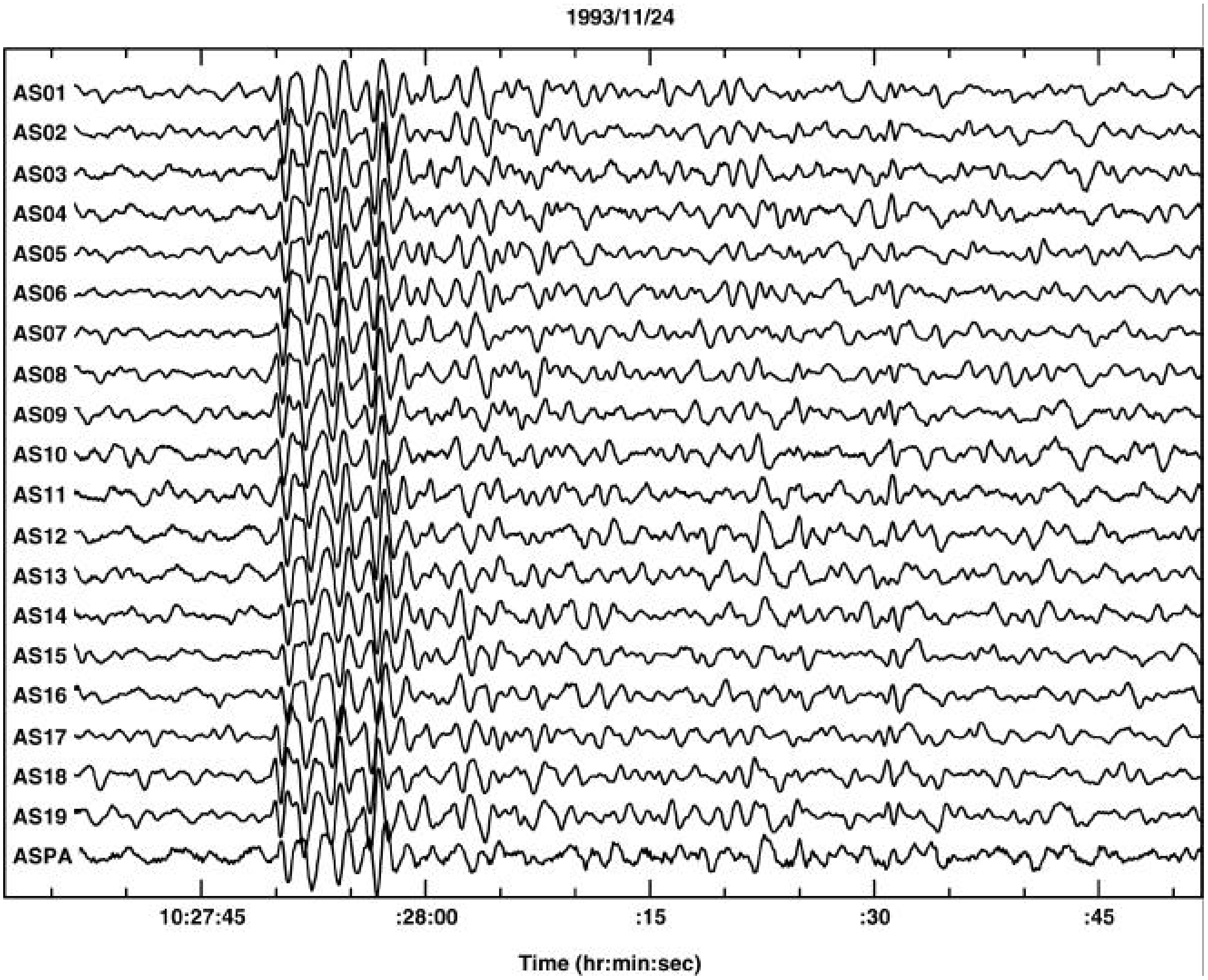

**Figure 5. Waveforms at the Alice Springs array for the November 1993 event. Vertical channels only.**

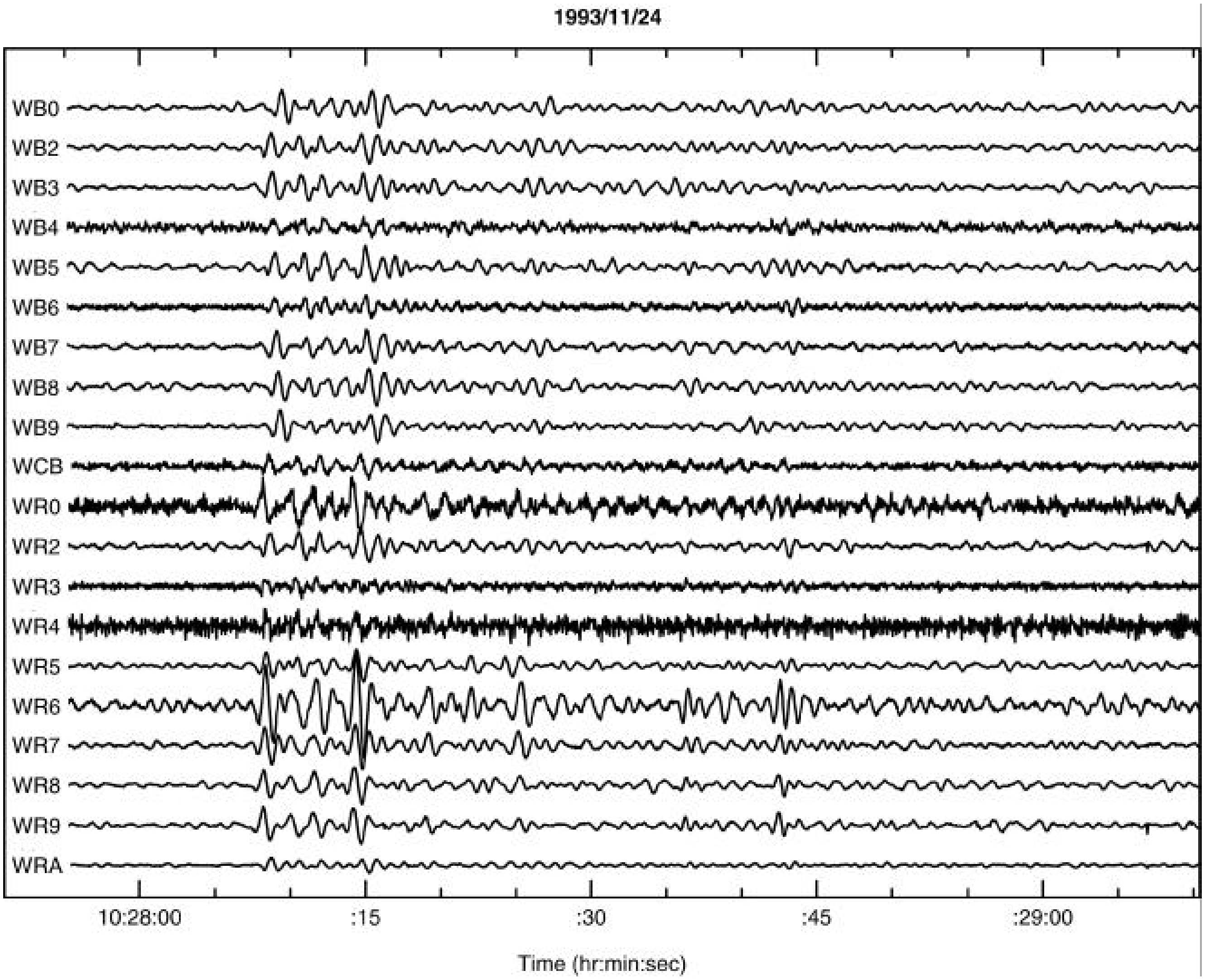

**Figure 6. Waveforms at the Warramunga array for the November 1993 event. Vertical channels only. These arrivals show the pattern of compressional wave energy discussed in the text.**

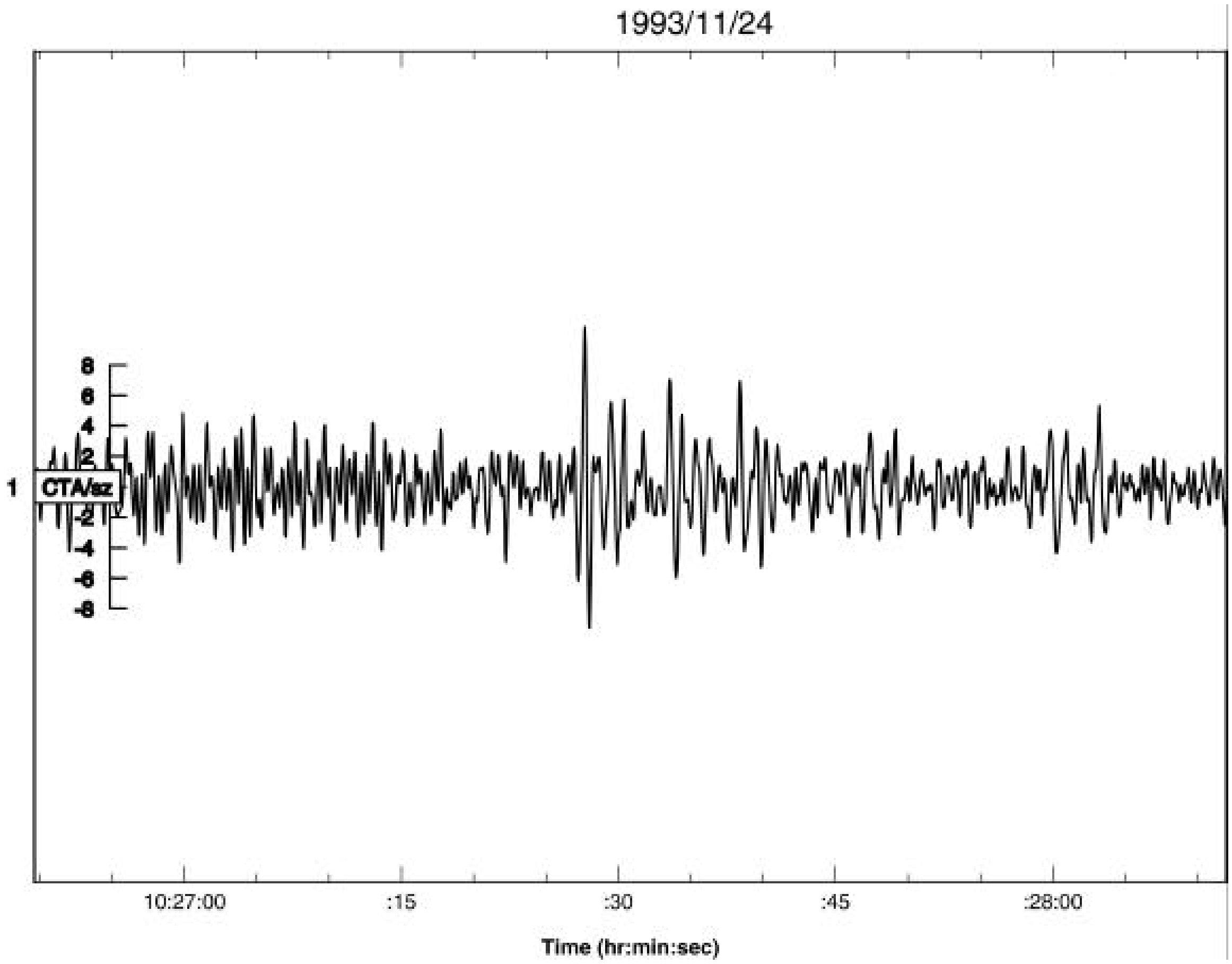

**Figure 7. Waveform from CTA for November 1993 event**

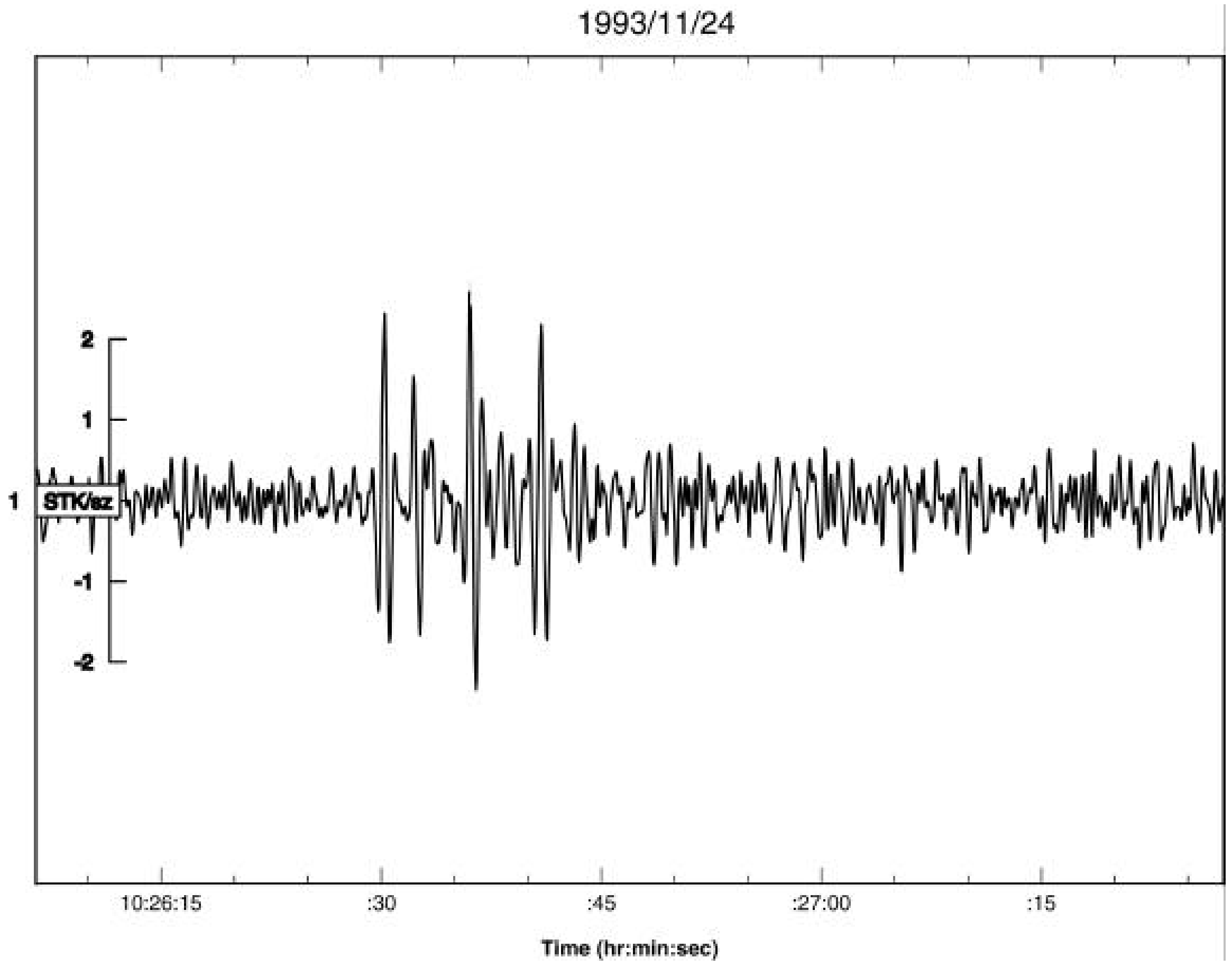

**Figure 8. Waveforms from STK for November 1993 event**

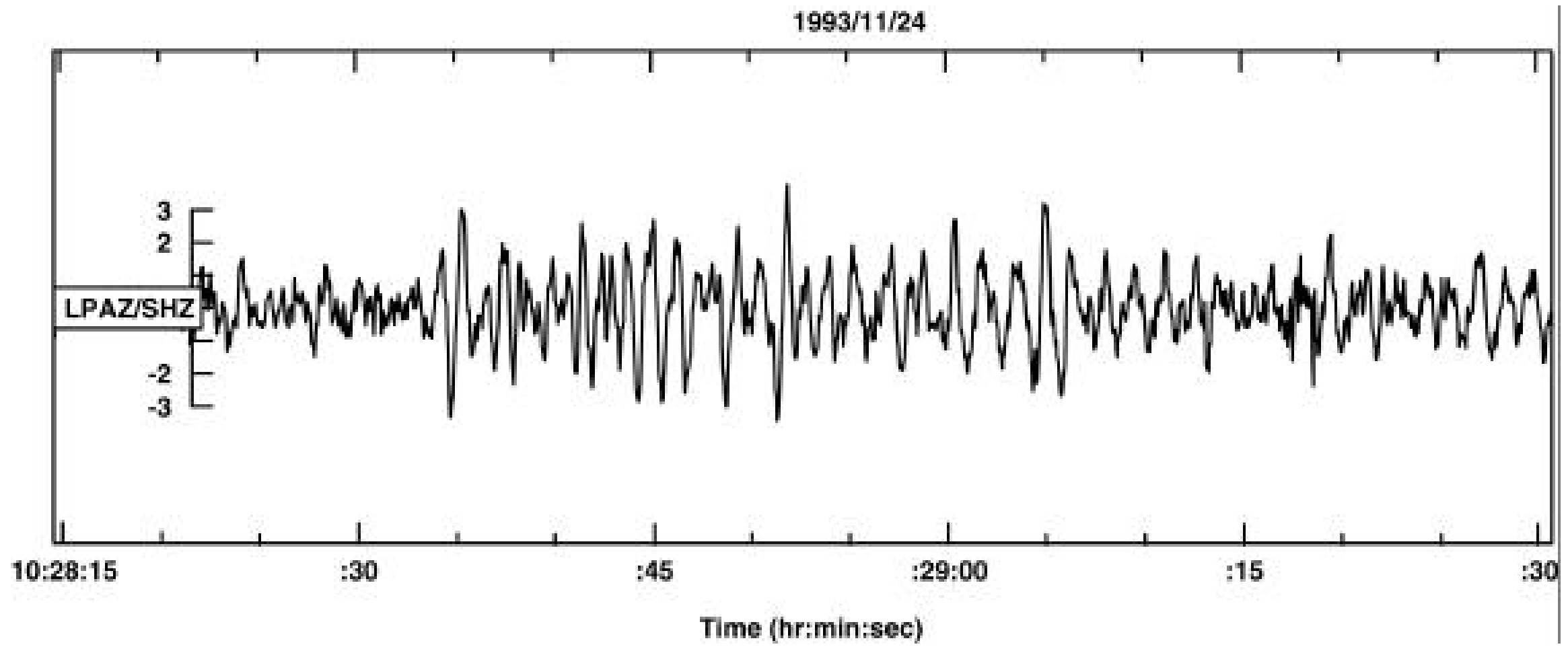

**Figure 9. Waveform on the vertical channel at La Paz for the November 1993 event.**

## VII. Summary and Conclusions

We have found 18 sets of 10 or more unassociated seismic reports from a set of 48 sensitive modern seismographic stations in 10 minute windows over a period during which Poisson statistics for random reports would predict we should have found one. Only one of these sets of reports is fit by a point source (or multiple point sources) model.

We have found one set (November, 1993) of 9 reports, 7 from members of the 48 "Class I" stations, that well fit the model of a line event that could be caused by passage through the Earth of a body of nuclear density and size of a few tens of microns. The fit to such a model from the pattern of arrival times is supplemented by data from two arrays consistent with initial signal origin at the point of closest approach (POCA) of the hypothetical linear trajectory and is also supported by waveforms available from 3 of the 9 stations involved that show the signal decreasing but not disappearing at the times predicted by the model.

Based on these results, we conclude that the dramatic excess of numbers of unassociated seismic reports over statistical expectation almost certainly indicates that these sets of reports are causally related and not random.  Their origin is an important subject for future seismological research.

The November, 1993, event has a natural explanation in terms of passage of a dense strange quark nugget through the Earth.

Acknowledgement:

The authors wish to thank Dr. James Taggart for making the USGS database available to us, Dr. Ken Muirhead who provided digital waveform data from the Australian stations, Dr. Thomas Coan, Dr. Doris Rosenbaum and Dr. Edward Wright for helpful discussions, and Oliver Morton and Tom Sigfried for valuable feedback.